\documentclass[allclo]{FBSart}
\usepackage{amsfonts}
\usepackage{amssymb}
\usepackage{amsmath}
\usepackage{graphicx}

\newcommand{\bbR}{\mathbb{R}}
\newcommand{\bbC}{{\mathbb C}}
\newcommand{\lal}{\langle}
\newcommand{\ral}{\rangle}
\newcommand{\bx}{\bm{x}}

\newcommand{\bk}{\bm{k}}
\newcommand{\wbk}{\widehat{\bm{k}}}
\newcommand{\bp}{\bm{p}}
\newcommand{\bq}{\bm{q}}
\newcommand{\wbq}{\widehat{\bm{q}}}
\newcommand{\ri}{\mathrm{i}}
\newcommand{\ra}{\mathrm{a}}
\newcommand{\rj}{\mathit{j}}
\newcommand{\rjd}{{\mathit{j}^\dagger}}

\title{Explicit Representations for the T-Matrix on Unphysical Energy
Sheets and Resonances in Two- and Three-Body Systems\thanks{Based on
a talk given at the International Workshop ``Critical Stability of
Few-Body Quantum Systems'' (Dresden, October 17--22, 2005). This
work was supported by the Deutsche Forschungsgemeinschaft (DFG) and
the Russian Foundation for Basic Research.}}
\author{Alexander K. Motovilov} \institute{Bogoliubov Laboratory of
Theoretical Physics, Joint Institute for Nuclear Research,
Joliot-Curie 6, 141980 Dubna, Moscow Region, Russia}

\runningauthor{A.\,K.\,Motovilov} \runningtitle{Explicit
Representations for the T-Matrix on Unphysical Sheets} 

\begin{document}

\maketitle
\begin{abstract}
We describe basic structure of the two- and three-body T-matrices,
scattering matrices, and resolvents continued to the unphysical
energy sheets. The description is based on the explicit
representations that have been found for analytically continued
kernels of the T-operators.
\end{abstract}

\section{Introduction}

Resonance is one of the most interesting and intriguing phenomena in
quantum scattering. With a resonance one usually associates an
unstable state that only exists during a certain time. The original
idea of interpreting resonances in quantum mechanics as complex
poles of the scattering amplitude (and hence, as those of the
scattering matrix) goes back to G.\,Gamov (1928). For radially
symmetric potentials, the interpretation of two-body resonances as
poles of the analytic continuation of the scattering matrix has been
entirely elaborated in terms of the Jost functions. Beginning with
E.\,C.\,Titchmarsh (1946) it was also realized that the s-matrix
resonances may show up as poles of the analytic continuation of the
Green functions.

Another, somewhat distinct approach to resonances is known as the
complex scaling (or complex rotation) method. The complex scaling
makes it possible to rotate the continuous spectrum of the $N$-body
Hamiltonian in such a way that resonances in certain sectors of the
complex energy plane turn into usual eigenvalues of the scaled
Hamiltonian. In physics literature the origins of such an approach
are traced back at least to C.\,Lovelace (1964). A rigorous approval
of the complex scaling method has been done by E.\,Balslev and
J.\,M.\,Combes (1971). A link between the s-matrix interpretation of
resonances and its complex rotation counterpart was established by
G.\,A.\,Hagedorn (1979) who has proven that for a wide class of
potentials the scaling resonances are also the scattering matrix
resonances.

If support of the interaction is compact, the resonances of a
two-body system can be treated within the approach created by
P.\,Lax and R.\,Phillips (1967). An advantage of the Lax-Phillips
approach is in the opportunity of giving an elegant operator
interpretation of resonances. The resonances show up as the discrete
spectrum of a dissipative operator which is the generator of the
compressed evolution semigroup.

For more details on the history of the subject and other approaches
to resonances, as well as for a bibliography we refer to paper
\cite{MN1997}. Here we only notice that, in contrast to the
``normal'' bound and scattering states, the resonant ones are still
a mysterious object and many questions related to resonances remain
unanswered. In particular, it is still unknown how to describe (in
an indisputable way) the scattering of a particle on a resonant
state of two other particles. To say the truth, there is a problem
even with definition of resonance since resonances are \emph{not} a
unitary invariant of a self-adjoint (Hermitian) operator. Moreover,
following to B.\,Simon \cite{Simon}, one may conclude that no
satisfactory definition can rely on a single Hamiltonian and always
an extra structure is necessary. Say, an unperturbed dynamics (in
quantum scattering theory) or geometric setup (in acoustical or
optical problems). Resonances are always relative as the scattering
matrix itself.

In our approach we follow the typical setup where the resonances
arising due to an interaction $V$ are considered relative to the
unperturbed dynamics described by the kinetic energy operator $H_0$.
The resolvent $G(z)=(H-z)^{-1}$ of the total Hamiltonian $H=H_0+V$
is an analytic operator-valued function of
$z\in\bbC\setminus\sigma(H)$. The spectrum $\sigma(H)$ of $H$ is a
natural boundary for holomorphy domain of $G(z)$ considered as an
operator-valued function.  However the kernel $G(\cdot,\cdot,z)$ may
admit analytic continuation through the continuous spectrum of $H$.
Or the form $\lal G(z)\varphi,\psi\ral$ may do this for any
$\varphi,\psi$ of a dense subset of the Hilbert space
$\mathfrak{H}$. Or the ``compressed'' resolvent $PG(z)P$ admits such
a continuation for $P$ the orthogonal projection onto a subspace of
$\mathfrak{H}$. In any of these cases one deals with the Riemann
surface of an analytical function.

In the simplest example with $H=H_0=-\Delta$, the two-body kinetic
energy operator in coordinate representation, we have
$$
G(\bx,\bx',z)=\frac{1}{4\pi}\frac{\mathrm{e}^{\ri
z^{1/2}|\bx-\bx'|}}{|\bx-\bx'|},
$$
where $\bx,\bx'$ are three-dimensional vectors. Clearly,
$G(\bx,\bx',z)$ as a function of the energy $z$ has a two-sheeted
Riemann surface which simply coincides with that of the function
$z^{1/2}$.

In this way one also arrives with the concept of the unphysical
energy sheet(s). The copy of the complex energy plane where the
resolvent $G(z)$ is considered initially as an operator-valued
function is called the physical sheet. The remainder of the Riemann
surface is assumed to consist of the unphysical sheets (in general,
an unphysical sheet may only be a small part of the complex plane).

Analytic function is uniquely defined by its values on an infinite
set in $\bbC$ having limiting point(s). Thus, if one knows the
resolvent ($T$-matrix, $S$-matrix) on the physical sheet then one
may, in principle, to express it on unphysical sheets through its
values in the physical sheet. In this case all the study of
resonances would reduce to a work completely on the physical sheet.

In \cite{MN1997,TMF1993} we have found just such expressions. More
precisely, we have derived explicit representations for the two- and
three-body $G(z)$, $T(z)$, and $S(z)$ on unphysical energy sheets in
terms of these quantities themselves only taken on the physical
sheet. In particular, the representations obtained show which blocks
of the scattering matrix are ``responsible'' for resonances on a
certain unphysical sheet.

\section{Two-body problem}

In general, we assume that the interaction potential $v$ falls off
in coordinate space not slower than exponentially. When studying
resonances of a two--body system with such an interaction one can
employ equally well both coordinate and momentum representations.
However in the three-body case it is much easier for us to work in
the momentum space (for an explanation see \cite{MN1997}, p. 149).
This is one of the reasons why we proceed in the same way in the
two-body case. Thus, for the two-body kinetic energy operator $h_0$
we set $(h_0 f)(\bk)=\bk^2 f(\bk)$ where $\bk\in\bbR^3$ stands for
the reduced relative momentum. In case of a local potential we have
$v(\bk,\bk')=v(\bk-\bk')$ and $v(\bk)=\overline{v(-\bk)}$. For
simplicity we assume that the function $v(\bk)$ is holomorphic in
$\bk$ on the whole three-dimensional complex space $\bbC^3$.

The transition operator (t-matrix) reads
\begin{equation}
\label{T} t(z)=v-vg(z)v,
\end{equation}
where $g(z)=(h-z)^{-1}$ denotes the resolvent of the perturbed
Hamiltonian $h=h_0+v$. The operator $t$ is the solution of the
Lippmann-Schwinger equation
\begin{equation}
\label{LScht2} t(z)=v-vg_0(z)t(z),
\end{equation}
that is,
\begin{equation}
\label{LSh1}
t(\bk,\bk',z)=v(\bk,\bk')-\int_{\bbR^3}d\bq
\dfrac{v(\bk,\bq)t(\bq,\bk',z)}{\bq^2-z}
\end{equation}
taking into account that
$g_0(\bk,\bk',z)={\delta(\bk-\bk')}/{(\bk^2-z)}$.

Clearly, all dependence of $t$ on $z$ is determined by the integral
term  on the r.h.s. part of \eqref{LSh1} that looks like a
particular case of the Cauchy type integral
\begin{equation}
\label{Phi} \Phi(z)=\int_{\bbR^N}d\bq\dfrac{f(\bq)}{\lambda+
\bq^2-z}
\end{equation}
for $N=3$. Cauchy integrals of the same form but for both $N=3$ and
$N=6$ we will also have below in three-body equations of Sec. 3.

Let $\mathfrak{R}_\lambda$, $\lambda\in\bbC$, be the Riemann surface
of the function
$$
\zeta(z)=\left\{\begin{array}{ll}
(z-\lambda)^{1/2}, & N \quad\text{odd}, \\
\log(z-\lambda), & N\quad\text{even}.
\end{array}
\right.
$$
If $N$ is odd, $\mathfrak{R}_\lambda$ is formed of two sheets of the
complex plane. One of them, where $(z-\lambda)^{1/2}$ coincides with
the arithmetic square root $\sqrt{z-\lambda}$, we denote by $\Pi_0$.
The other one, where $(z-\lambda)^{1/2}=-\sqrt{z-\lambda}$, is
denoted by $\Pi_1$.

If $N$ is even, the number of sheets of $\mathfrak{R}_\lambda$ is
infinite. In this case as the index $\ell$ of a sheet $\Pi_\ell$ we
take the branch number of the function $\log(z-\lambda)$ picked up
from the representation
$\log(z-\lambda)=\log|z-\lambda|+\ri\,2\pi\ell +\ri\phi$ with
$\phi\in[0,2\pi)$.

The following statement can be easily proven by applying the residue
theorem (if necessary, see \cite{TMF1993} for a proof).
\smallskip

\noindent{\bf Lemma 1}. {\it For a holomorphic $f(\bq)$,
$\bq\in\bbC^N$, the function $\Phi(z)$ given by \eqref{Phi} is
holomorphic on $\bbC\setminus[\lambda, +\infty)$ and admits the
analytic continuation onto $\mathfrak{R}_\lambda$ as follows
\begin{equation}
\label{Phi1} \Phi(z)|_{\Pi_{\ell}}=\Phi(z)-\ell\,\pi\ri
(\sqrt{z-\lambda}\,\,)^{N-2} \int_{S^{N-1}}d\widehat{\bq}\,
f(\sqrt{z-\lambda}\wbq),
\end{equation}
where $S^{N-1}$ denotes the unit sphere in $\bbR^N$ centered at the
origin. (Position of the argument $z$ in the sheet $\Pi_0$ on the
r.h.s. part of \eqref{Phi1} is the same as that of the argument of
$\Phi|_{\Pi_{\ell}}$ on the sheet $\Pi_{\ell}$.)}

Now set $ \bigl(g_0(z)f_1,f_2\bigr)\equiv
\int_{\bbR^3}d\bq\,\frac{f_1(\bq)f_2(\bq)}{\bq^2-z}$ where $f_1$ and
$f_2$ are holomor\-phic. Then by Lemma 1
$$
\bigl(g_0(z)f_1,f_2\bigr)|_{\Pi_1}=\bigl(g_0(z)f_1,f_2\bigr)|_{\Pi_0}-
\pi\ri\sqrt{z}\int_{S^{2}}d\widehat{\bq}\,
f_1(\sqrt{z}\wbq)f_2(\sqrt{z}\wbq).
$$
which means that the continuation of the free Green function
$g_0(z)$ onto the unphysical sheet $\Pi_1$ can be written in short
form as
\begin{equation}
\label{g0c} g_0(z)|_{\Pi_1}=g_0(z)+\ra_0(z)\rj^\dagger(z)\rj(z),
\end{equation}
where $\ra_0(z)=-\pi\ri\sqrt{z}$ and $\rj(z)$ is the operator
forcing a (holomorphic) function $f$ to set onto the energy shell,
i.e. $\bigl(j(z)f\bigr)({\wbk})=f(\sqrt{z}\wbk)$.

Taking into account \eqref{g0c}, on the unphysical sheet $\Pi_1$ the
Lippmann-Schwinger equation \eqref{LScht2} turns into
$$
t'=v-v(g_0+\ra_0\rjd\rj) t', \qquad t'=t|_{\Pi_1}.
$$
Hence $ (I+vg_0)t'=v-\ra_0\rjd\rj\, t'.$ Invert $I+vg_0$ by using
the fact that $t(z)=v-vg_0t$ and, hence, $(I+vg_0)^{-1}v=t$:
\begin{equation}
\label{t1} t'=t-\ra_0 t\rjd\rj t'.
\end{equation}
Apply $\rj(z)$ to both parts of \eqref{t1} and obtain $\rj t'=\rj
t-\ra_0\, \rj t\rjd\,\rj t'$,  which means
\begin{equation}
\label{t11} (I+\ra_0\, \rj t\rjd)\,\rj t'=\rj t.
\end{equation}
Observe that $I+\ra_0\, \rj t\rjd$ is nothing but the scattering
matrix $s(z)$ since the kernel of $s(z)$ reads
$$
s(\wbk,\wbk',z)=\delta(\wbk,\wbk')-\pi\ri\sqrt{z}
\,\,t(\sqrt{z}\wbk,\sqrt{z}\wbk',z).
$$
Hence $\rj t'=[s(z)]^{-1}\rj t$. Now go back to \eqref{t1} and get
$t'=t-\ra_0 \,\,t\rjd [s(z)]^{-1} \rj t$, that is,
\begin{equation}
\label{t11p}  t(z)|_{\Pi_1}=t(z)-\ra_0(z)
\,\,t(z)\rjd(z)[s(z)]^{-1}\,\rj(z)t(z).
\end{equation}
All entries on the r.h.s. part of \eqref{t11p} are on the physical
sheet. This is just the representation for the t-matrix on the
unphysical sheet we looked for.

{}From \eqref{t11p} one immediately derives representations for the
continued resolvent,%
\begin{equation}
\label{gr}
 g(z)|_{\Pi_1}=g+\ra_0\,(I-gv)\rjd\,[s(z)]^{-1}\rj(I-vg),
\end{equation}
and continued scattering matrix,
\begin{equation}
\label{sr} s(z)|_{\Pi_1}=\mathcal{E}\,[s(z)]^{-1}\,\mathcal{E},
\end{equation}
where $\mathcal{E}$ is the inversion,
$(\mathcal{E}f)(\wbk)=f(-\wbk)$. Hence, the resonances are nothing
but zeros of the scattering matrix $s(z)$ in the physical sheet.
That is, the energy $z$ on the unphysical sheet $\Pi_1$ is a
resonance if and only if there is a non-zero vector $\mathcal{A}$ of
$L_2(S^2)$ such that $s(z)\mathcal{A}=0$ for the same $z$ on the
physical sheet.

The function $\mathcal{A}(\wbk)$ is the breakup amplitude of the
resonance state.  This means that  in coordinate space the
corresponding ``Gamov vector'' (the resonance solution to the
Schr\"odinger equation) has the following asymptotics
$$
\psi_{\mathrm{res}}(\bx)\mathop{\sim}\limits_{\bx\to\infty}
\mathcal{A}(-\widehat{\bx})\dfrac{\mathrm{e}^{-\ri\sqrt{z}|\bx|}}{|\bx|}.
$$

\section{Three-body problem}

Let $H_0$ be the three-body kinetic energy operator in the
center-of-mass system. Assume for simplicity that there are no
three-body forces and thus the total interaction reads
$V=v_1+v_2+v_3$ where $v_\alpha$, $\alpha=1,2,3$, are the
corresponding two-body potentials having just the same properties as
in the previous section.

The best way to proceed in the three-body case is to work with the
Faddeev components \cite{Faddeev1963}
$$
M_{\alpha\beta}=\delta_{\alpha\beta}v_{\alpha}-v_{\alpha}
G(z)v_{\beta} \quad (\alpha,\beta=1,2,3)
$$
of the T-operator $T(z)=V-VG(z)V$ where, as usually, $G(z)$ denotes
the resolvent of the total Hamiltonian $H=H_0+V$. The components
$M_{\alpha\beta}$ satisfy the Faddeev equations
\begin{equation}
\label{F}
M_{\alpha\beta}(z)=\delta_{\alpha\beta}\mathbf{t}_\alpha(z)-
\mathbf{t}_\alpha(z) G_0(z)\sum_{\gamma\neq\alpha}M_{\gamma\beta}(z)
\end{equation}
with $G_0(z)=(H_0-z)^{-1}$ and
$\mathbf{t}_\alpha(P,P',z)=t_\alpha(\bk_\alpha,\bk'_\alpha,z-\bp_\alpha^2)
\delta(\bp_\alpha-\bp'_\alpha)$ where $\bk_\alpha,\bp_\alpha$ denote
the corresponding reduced Jacobi momenta (see \cite{MN1997} for the
precise definition we use) and $P=(\bk_\alpha,\bp_\alpha)\in\bbR^6$
is the total momentum.

Assume that any of the two-body subsystems has only one bound state
with the corresponding energy $\varepsilon_\alpha<0$,
$\alpha=1,2,3$. Assume in addition that all of these three binding
energies are different. It is easy to see that the thresholds
$\varepsilon_1$, $\varepsilon_2$, $\varepsilon_3$, and 0 are
associated with particular Cauchy type integrals in the integral
equations \eqref{F}. By Lemma 1 the two-body thresholds
$\varepsilon_\alpha$ appear to be square-root branching points while
the three-body threshold 0 is the logarithmic one. In order to
enumerate the unphysical sheets we introduce the multi-index
$\ell=(\ell_0,\ell_1,\ell_2,\ell_3)$ with
$\ell_0=\ldots,-1,0,1,\ldots$ and $\ell_\alpha=0,1$ if
$\alpha=1,2,3$. Clearly, only encircling the two-body thresholds one
arrives at seven unphysical sheets. The three-body threshold
generates infinitely many unphysical sheets. (There might also be
additional branching points on the unphysical sheets, in particular
due to two-body resonances.)

It turns out that the analytically continued equations \eqref{F} can
be explicitly~(!) solved in terms of the matrix
$M=\{M_{\alpha\beta}\}$ itself taken only on the physical sheet,
just like in the case of the two-body t-matrix. But, of course, now
the result depends on the unphysical sheet $\Pi_{\ell}$ concerned.
More precisely, the resulting representation reads as follows
\begin{equation}
\label{MPi} M|_{\Pi_{\ell}}=M+Q_M
\,L\,S_{\ell}^{-1}\,\widetilde{L}\,\widetilde{Q}_M.
\end{equation}
In the particular case we deal with, $L$ and $\widetilde{L}$ are
$4\times4$ scalar matrices of the form
$L=\mathop{\mathrm{diag}}({\ell}_0, {\ell}_1,$ ${\ell}_2,{\ell}_3)$
and
$\widetilde{L}=\mathop{\mathrm{diag}}
(|{\ell}_0|,{\ell}_1,{\ell}_2,{\ell}_3)$,
respectively; $S_{\ell}(z)=I+L(S(z)-I)\widetilde{L}$ is a truncation
of the total scattering matrix $S(z)$ and the entries $Q_M$,
$\widetilde{Q}_M$ are explicitly written in terms of the
half-on-shell kernels of $M$ (see formula (7.34) of \cite{MN1997}).
{}From \eqref{MPi} one also derives explicit representations for
$G(z)|_{\Pi_{\ell}}$ and $S(z)|_{\Pi_{\ell}}$ similar to those of
\eqref{gr} and \eqref{sr}, respectively.

Thus, to find resonances on the sheet $\Pi_{\ell}$ one should simply
look for the zeros of the truncated scattering matrix $S_{\ell}(z)$,
that is, for the points $z$ in the physical sheet where equation
$S_{\ell}(z)\mathcal{A}=0$ has a non-trivial solution $\mathcal{A}$.
The vector $\mathcal{A}$ will consist of breakup amplitudes of the
resonance state into the various possible channels. Within such an
approach one can also find the virtual states.

In order to find the amplitudes involved in $S_{\ell}$, one may
employ any suitable method, for example the one of Refs.
\cite{CPC2000,YaF1999,YaF1997} based on the Faddeev differential
equations. In these works the approach we discuss has been
successfully applied to several three-body systems. In particular,
the mechanism of emerging the Efimov states in the $^4$He trimer has
been studied \cite{CPC2000,YaF1999}.

\begin{acknowledge}
The author is grateful to Prof. Dr. Sergio Albeverio for his
hospitality at the Institute for Applied Mathematics, University of
Bonn, where this paper has been completed.
\end{acknowledge}

\end{document}